\documentclass[runningheads]{llncs}
\usepackage{graphicx}
\usepackage{booktabs,subcaption,amsfonts,dcolumn,graphicx,xcolor,amsmath}
\usepackage{xcolor}
\usepackage{multirow}
\usepackage{hhline}
\usepackage{array}
\usepackage{tabu}
\usepackage{makecell}
\usepackage{booktabs}
\pdfoutput=1

\begin{document}
%
% \title{ThreatZoom: Hierarchical Neural Network for CVEs to CWEs Classification}
% \title{ThreatZoom: Hierarchical Neural Network for CVEs to CWEs Classification}
\title{ThreatZoom: CVE2CWE using Hierarchical Neural Network}
% \titlerunning{Automatic Classification of CVEs to CWEs}
% \author{
\author{Ehsan Aghaei
\and
Waseem Shadid \and
Ehab Al-Shaer
}
% \authorrunning{Anonymous}
%
\authorrunning{Aghaei et al.}
% \institute{\\
% \email{}}
\institute{University of North Carolina at Charlotte, Charlotte, USA, 28262 \\
\email{\{eaghaei, wshadid, ealshaer\}@uncc.edu}}
\maketitle              % typeset the header of the contribution
\begin{abstract}
The Common Vulnerabilities and Exposures (CVE) represent standard means for sharing publicly known information security vulnerabilities. One or more CVEs are grouped into the Common Weakness Enumeration (CWE) classes for the purpose of understanding the software or configuration flaws and potential impacts enabled by these vulnerabilities and identifying means to detect or prevent exploitation.\\

As the CVE-to-CWE classification is mostly performed manually by domain experts, thousands of critical and new CVEs remain unclassified, yet they are unpatchable. This significantly limits the utility of CVEs and slows down proactive threat mitigation.
This paper presents the first automatic tool to classify CVEs to CWEs. {\em ThreatZoom} uses a novel learning algorithm that employs an adaptive hierarchical neural network which adjusts its weights based on text analytic scores and classification errors. It automatically estimates the CWE classes corresponding to a CVE instance using both statistical and semantic features extracted from the description of a CVE. 
\\

This tool is rigorously tested by various datasets provided by MITRE and the National Vulnerability Database (NVD). The accuracy of classifying CVE instances to their correct CWE classes is $92\%$ (fine-grain) and $94\%$ (coarse-grain) for NVD dataset, and $75\%$ (fine-grain) and $90\%$ (coarse-grain) for MITRE dataset, despite the small corpus. 

%  However, major challenges including extracting the appropriate features, considering the CVE-CWE semantic gap, and hierarchical fine-grain classification to multiple CWE classes need to be addressed.
\end{abstract}
\section{Introduction}
%What is the problem and why it is important 
% [*************************************\]

Cyber-attack actions allow malicious actors to violate the intended security policies by bypassing the protection mechanisms or manipulating the resources or the system's behaviors. Thus, the consequence of the attack is a behavior that violates the intended security policies of the victim. Such action, if it is made accessible to the attacker, is called weakness. Weaknesses are flaws, fault, bugs, and errors occur in software's architecture, design, code, or implementation that can lead to exploitable vulnerabilities. A vulnerability is defined as a set of one or more weaknesses within a specific product or protocol, which allows an attacker to access the behaviors or resources to compromise.

The Common Weakness Enumeration (CWE) is a hierarchically-designed dictionary of software weaknesses for the purpose of understanding software flaws, their potential impacts, and identifying means to detect, fix, and prevent errors \cite{(3)CweMITRE2018}. 
CWEs are non-disjoint classes, and they are organized in a hierarchical (tree) structure to reflect this relationship in which every non-root CWE node inherits the whole characteristics of its parents. Therefore, a CWE in the higher level represents a more general definition of weakness, and every lower-level node adds more details to that CWE node of interest.
Meanwhile, the Common Vulnerabilities and Exposures (CVE) reports are a list of publicly disclosed computer security vulnerabilities where every report is assigned by an ID. CVEs are brief and low-level descriptions, representing the standard means for sharing cybersecurity vulnerabilities within a particular product or system  \cite{(2)CveMITRE2018}. CVE IDs are assigned to those vulnerabilities that satisfy three criteria: (1) system bugs and its negative impact on security must be acknowledged by the vendor, or vulnerability report and security policy violation of the affected system must be documented by the reporter, (2) bug can be fixed independently of any other bugs, (3) bug must affect only one codebase and those impact more than one system get separate IDs.

In general, CWE is a dictionary of software vulnerabilities addressing the underlying error, while CVE is a set of known instances of vulnerability for specific systems or products. 
CWEs mainly explain \textit{how} (conditions and procedures) and \textit{why} (cause) a vulnerability can be exploited, and \textit{what} (impact) are the consequences. When the vulnerability is unpatchable, the classification of CVE into CWE becomes extremely important since only CWEs can provide the means to develop countermeasure techniques and understand the CVE implications. In this paper, we present a novel approach that employs a hierarchical neural network design to automatically estimate the  CWE classes corresponding to a CVE instance using both statistical and semantic features from the description of CVEs.

\subsection{Motivation Example}
Given 'CVE-2004-0366: \textit{SQL injection vulnerability in the libpam-pgsql library before 0.5.2 allows attackers to execute arbitrary SQL statements.}', the description shares the attack action (execute arbitrary SQL statement) within a particular object (libpam-pgsql library) and specifies the consequence (SQL injection). 
Although this low-level and product-oriented definition demonstrates the exploitation of SQL injection, it fails to adequately specify the characteristic of this malicious behavior, which is necessary to address potential prevention and/or detection methods.
The complementary associated CWE (CWE-89: SQL Injection) \footnote{https://cwe.mitre.org/data/definitions/89.html} provides a high-level and beyond-the-product knowledge by answering three key questions: (1) why the attack is exploited: {\em the system does not or incorrectly neutralized special elements}, (2) how this is exploited:  {\em by modifying the intended SQL command}, and (3) what the possible consequences are: {\em read or modify application data}, and {\em bypass protection mechanism}. 

The above-mentioned case is a confirmatory example to show how a CWE can paint a clear picture of the existing holes in the systems and reveals potential factors that cause vulnerability exploitation. Obtaining these factors is closely associated with the paradigm of pinpointing applicable mitigation or detection methods. For example, we can apply an "accept known good" input validation strategy, i.e., using a set of legit inputs that strictly conform to specifications and rejects the rest, to mitigate SQL injection. Besides, we can detect SQL injection by performing an automated static analysis (e.g., bytecode or binary weakness analysis), dynamic analysis (e.g., database or web service scanners), or design review (e.g., formal methods).
\begin{figure}
\centerline{\includegraphics [width=1\textwidth]{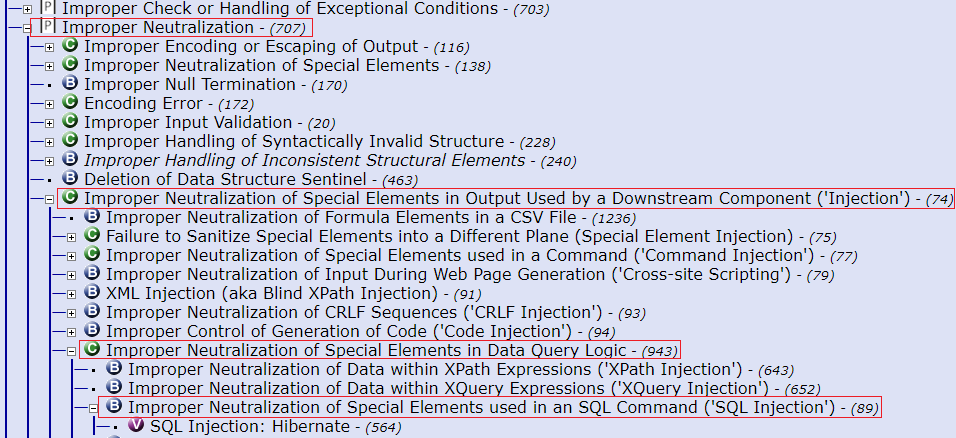}}
\caption {It depicts the hierarchical representation of the CWEs. The red boxes show the CWE-89's relatives in the higher levels. This hierarchy plays an important role in understanding the character of the weaknesses from different levels.}
\label{fig:cwesnap}
\end{figure}
\vspace{-0.03in}
Fig. \ref{fig:cwesnap} shows the hierarchical representation of the CWEs. Analyzing the path from the root all the way to any node in the lower levels is indispensable since each node reveals different functional directions to learn a weakness. For example, by tracking the path from the root node, CWE-707, to  CWE-89, we realize that the SQL injection (CWE-89) is a result of an improper neutralization of special elements in data query logic (CWE-943), where both weakness are associated with injection (CWE-74), and the injection itself is the result of improper formation and neutralization of a message within a product before it is read from an upstream component or sent to a downstream component (CWE-707). Incorporating this thorough knowledge graph helps to maintain countermeasures from different senses, even if the most fine-grain node is not available. For example, assume that only two coarse-grain candidates in different levels of hierarchy, CWE-707, and CWE-74, are available for CVE-2004-0366, while the most fine-grain weakness (CWE-89) is not discovered yet. Although fine-grain SQL injection characteristics is not exposed, investigating the coarse-grain candidates helps to find the common consequences and impacts, and accordingly extract defense actions against improper neutralization and injection (e.g., filtering control-plane syntax from all input). This example explicitly highlights the significance of the existing hierarchical structure of CWEs and shows how useful it is in perceiving the defense actions.

\subsection{Related Works}
AA great effort has been made initially by MITRE and NVD\footnote{National Vulnerability Database} to manually classify some CVEs to one or more CWE classes. Considering the growing number of CVEs and the high labor cost of manual classification, MITRE has classified 2553 CVEs (out of ~116K) to 364 CWE classes (out of 719) \cite{(3)CweMITRE2018}. On the other hand, NVD has \cite{(9)NVD2018} attempted to increase the quantity of CVE classification by mapping about 85,000 CVEs. Although MITRE has classified smaller number of CVEs compared with NVD, it covers a higher number of CWEs, and performs a more fine-grain classification in  hierarchical fashion. In the meantime, NVD classified more CVEs but it includes smaller number of CWEs, without addressing the hierarchy. The further investigation about the characteristics of each effort is discussed in section \ref{section:datset}.

In the meantime, there have been several research efforts other than MITRE and NVD to analyze CVEs to enhance the searching process and to perform CVE categorization. Neuhaus \textit{et al.}~\cite{(22)Neuhaus2010} proposed a semi-automatic method to analyze the CVE descriptions using topic models to find prevalent vulnerability types and new trends. The test result reports 28 topics in these entries using Latent Dirichlet Allocation (LDA) and assigned LDA topics to CWEs. This approach shows very limited accuracy, depending on the CWE type.

Na \textit{et al.}~\cite{(23)Na2016} proposed  Na\"ive Bayes classifier to categorize CVE entries into the top ten most frequently used CWEs with the accuracy of 75.5\%. However, the accuracy of this limited classification significantly decreases as the number of the considered CWEs increases (i.e., accuracy decreased from 99.8\% to 75.5\% when the number of CWE classes increased from 2 to 10). In addition, this approach does not consider the hierarchical structure for CWEs, which significantly limits its value.
Another classifier was developed to estimate the vulnerabilities in CVEs using the basis of previously identified ones by Rahman \textit{et al.} \cite{(24)Rehman2012}. This approach uses the Term Frequency-Inverse Document Frequency (TF-IDF) to assign weights to text tokens form the feature vector and Support Vector Machine (SVM) to map CVEs to CWEs. However, they use only six CWE classes and 427 CVE instances. In addition, their classifier does not follow the hierarchical structure for CWEs as well.

\subsection{Challenges}
Our investigation has revealed there are three main challenges to solve the CVE-to-CWE classification as follows:

First, CWEs are organized in a hierarchical (tree) structure, in which, a CVE instance can be classified into one or more interdependent CWEs that belong to the same path. In the meantime, a CWE class can be accessed from more than one path in the CWE tree structure \cite{(3)CweMITRE2018}. For instance, CWE-22 (Path Traversal)\footnote{https://cwe.mitre.org/data/definitions/22.html} can be reached from two different paths, started by either CWE-435 (Improper Interaction Between Multiple Correctly-Behaving Entities) \footnote{https://cwe.mitre.org/data/definitions/435.html} or CWE-664 (Improper Control of a Resource Through its Lifetime) \footnote{https://cwe.mitre.org/data/definitions/664.html}.

Second, there is a semantic gap in and between the language of CVEs and CWEs that makes the feature extraction phase challenging. Thus, performing an efficient semantic analysis is inevitable in order to identify the connection between similar concepts used in the CVE and CWE context.

Third, a considerably small percentage (about $2\%$) but high-quality classification (fine-grains in the CWE hierarchy) of CVEs are provided by MITRE. On the other hand, NVD delivers a higher percentage of CVE classification ($71\%$), but it used a considerably lower portion of CWEs (about $32\%$ of CWEs used by MITRE). Hence, there should be a trade-off to process small and imbalanced datasets and classify CVEs into the most fine-grain CWE along with exploring the entire nodes in the path. In addition, the feature extraction, feature selection, and training process must be robust to handle the overfitting problem as well.
%there are about $50\%$ of the created CWE classes were not used in CVE classification. 

\subsection{Contribution}
Our research aims at finding a solution for this challenging problem by offering a novel automated system, so-called ThreatZoom, to estimate the CWE classes corresponding to a CVE instance. ThreatZoom takes the CVE's description as input and assigns a list of CWE classes along with the path connecting the roots of the tree to the lowest possible level class. This classification procedure comprises three steps: preprocessing (section \ref{sub:Preprocessing}), feature extraction (section \ref{sub:feature_extraction}) , and hierarchical
decision-making process (section \ref{sub:hir_dec}). 
The feature extraction algorithm extracts the textual features from the name and the description of all existing CWE classes and their associated CVEs.
Leveraging synonym vector coding and n-gram analysis, we extract both statistical and semantic features and process them to create feature vectors. Each feature is assigned by a TF-IDF score that reflects its importance in the feature space.

 The main novelty of this framework is in how TF-IDF scores are incorporated into the neural network design. TF-IDF scores are set to be the initial weights of the neural network at each level such that, the weights of the neural network at one level is the sum of all TF-IDF scores found in its children CWE classes. Then, the neural network framework is trained to adjust these scores to reduce the classification error during backpropagation. This unique technique of using TF-IDF scores within a neural network framework has three advantages: (1) it allows the neural network to learn experts perspectives in classifying CVE instances using his knowledge, and the provided descriptions about the CVE, (2) helps in reducing the effect of having a small number of training samples, and (3) increases the chances for the neural network to converge at the minimum point, which is highly likely is close to the global one in this classification problem. 

In summary, this work has four key contributions: (1) the development of the first algorithm that automatically classifies CVEs to CWE classes, (2) a novel design of a hierarchical neural network that is able to trace CWE classes to the most fine-grain in the CWE hierarchical structure, 
(3) a new approach to extract semantic and statistical features from CVE descriptions and compute the score that reflects the importance of a feature in describing the vulnerabilities at each level in the CWE hierarchical structure, 
and (4) an adaptive learning technique that incorporates the computed feature scores by TF-IDF in a neural network framework to allow an automatic adjustment for feature scores to reduce classification errors. The algorithm is tested using CVE instances provided by MITRE and NVD. The test results show high accuracy classification and allow for heightening threat understanding of CVE instances to take practical mitigation actions.

To the best of our knowledge, this presented solution is first to offer an automated fine-grain classification of CVE instances to their corresponding CWE classes using a large number of CWE classes and following its hierarchical structure.
\section{Methodology}
\label{sec:Methodology}
This section describes the algorithm to estimate the CWE classes associated with CVE instances. The algorithm takes as input a CVE instance, and produces a set of plausible CWE classes that might be associated with the input CVE.
\begin{figure}
\begin{centering}
\includegraphics[width = 12cm]{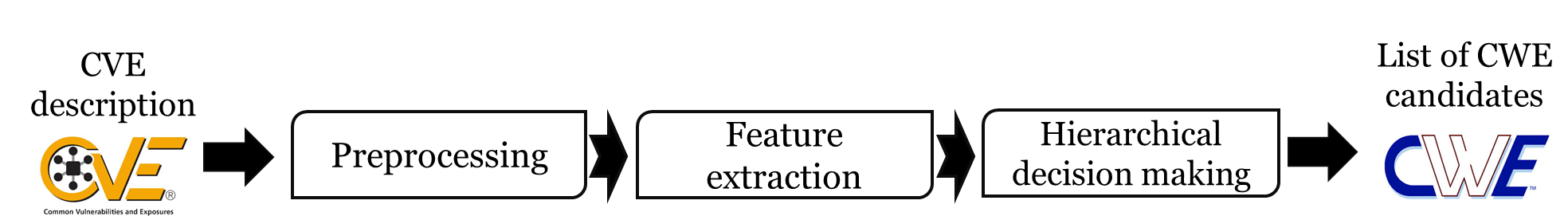}\caption{It shows the three steps for the algorithm to generate a list of plausible
CWE classes corresponding to the input CVE instance.\label{fig:3-1}}
\par\end{centering}
\end{figure}
The algorithm consists of three steps namely preprocessing, feature extraction, and hierarchical neural network decision-making process (Fig. \ref{fig:3-1}). 
The algorithm extracts the text from the description of input CVE instance, and cleans it to generate a normalized version of it. This normalized text is processed to extract n-gram features to characterize  the statistical and semantic properties. These features are passed to a trained hierarchical neural network framework to estimate the CWE classes associated with the input CVE. In the following subsections, we describe the details of each three steps in details.

\subsection{\label{sub:Preprocessing} Preprocessing}
This step takes the text provided in the description field of a CVE instance and turns it into a clean and normalized version. This preprocessing phase is accomplished via a five-stage process:

I. \textbf{Converting to lowercase}: Convert every letter to its lowercase. This is based on the assumption that  each letter conveys the same meaning regardless of its case or encoding.

II. \textbf{Filtering the stop words}: Remove words that do not convey valuable information for the classification process. These words are defined in the stopword list provided by Stanford Natural Language Processing Toolkit (NLTK), e.g., "to", "the", "for", "in".
    
III. \textbf{Text cleaning}: Remove punctuation and special characters that do not convey valuable information (e.g., ",", "!", "?"). Hyphens are kept though, because they may convey information.

IV. \textbf{Word stemming}: Reduce each word to its root in order to identify the relationships and commonalities across large text documents. All words are stemmed using snowball stemming model provided by NLTK python library.

V. \textbf{Synonym vector coding}: Groups the synonym words in the context, assigns a code word to represent each group, and replace all the synonym words with the code word which represents that group of synonym words in the context. MITRE provides a section for a portion of CWEs called "Alternative Terms" in which it provides the abbreviations or other commonly used advanced terms for that particular weakness. For example, [XEE, XML entity expansion, Billion Laughs Attack, XML Bomb] is a word vector associated with CWE-776 ('XML Entity Expansion'). In addition, MITRE provides a "CWE Glossary"\footnote{https://cwe.mitre.org/documents/glossary/}, which contains more general terminology that is interchangeably used in CWE descriptions to convey similar meanings. Each group of words is represented by a code that is a word from the group, such that any other word in the group will be replaced by this code if found in the text. For example, the word \textit{improper} shares the same meaning with \textit{insufficient} and \textit{incorrect} in cybersecurity domain, therefore they belong to the same vector and represented by code \textit{incorrect}. This helps in reducing the variance within the documents in the same category, and in increasing the distance between different categories.

By conducting this five-stage process, the text is clean and has a unified terminology. This helps in reducing the feature space dimension as described in the next section.

\begin{figure}
\begin{centering}
\includegraphics[width=12cm]{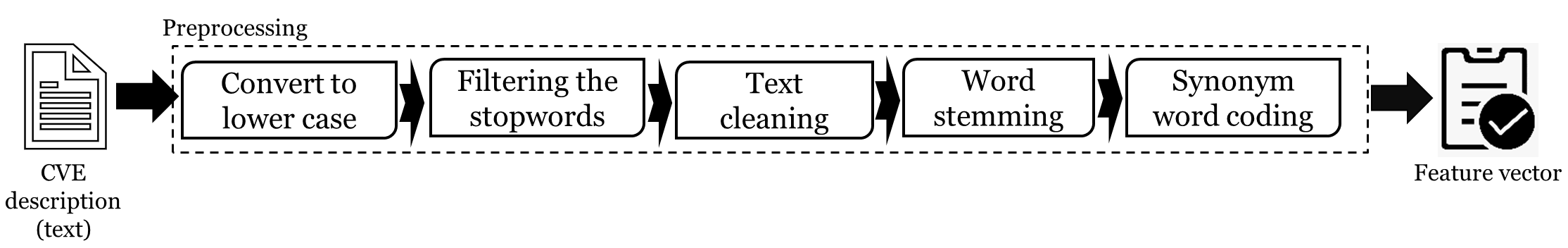}
\par\end{centering}
\caption{This shows the five CVE description preprocessing stages. 
%: lower-case conversion, stop-word
%filtration, text cleaning, stemming words. By performing these stages
%a clean, normalized and representative text is generated.
\label{fig:1} }
\end{figure}
\subsection{\label{sub:feature_extraction}Feature Extraction}
The input to this component is the preprocessed text and the output is word-level n-grams feature vectors \cite{(5)Ogada2015,aghaei2019host,aghaei2017ensemble}.
The feature extraction is carried out in two stages as follows:

% \begin{itemize}
\subsubsection{N-gram analysis:}
In this phase, the input text is analyzed to generate a list of unique 1-gram, 2-gram, and 3-gram terms.
The window size $n$ in the n-gram analysis is a hyperparameter and set experimentally. We tend to set this value to $1,2,3$ to satisfy stability and to avoid overfitting. Longer n-grams dramatically increase the number of terms, i.e., word combination, that are rarely found in CVE instances and may cause over-fitting in the training process.
The list of all unique n-gram items, $\mathbf{N}$, is the union of
$N_{1}$, $N_{2}$, $N_{3}$, i.e.: 

\[\mathbb{\mathbf{N}}= \{N_{1}\}\cup\{N_{2}\}\cup\{N_{3}\}\]
% $N_{1}$, $N_{2}$, $N_{3}$, i.e., $\mathbb{\mathbf{N}}= \{N_{1}\}\cup\{N_{2}\}\cup\{N_{3}\}$.
This list of unique n-gram terms may contain terms that are rarely found in describing CVE instances or CWE classes. Hence, these terms are removed from the list to generate the feature vector as described next.

\subsubsection{Feature vector generation:} 
This stage takes the list of unique n-gram terms extracted in the first stage as input and generates a multi-hot representation vector indicating the terms that are likely to be used in describing CVE instances and CWE classes. This vector is referred to as a feature vector.
The collection of terms that are likely to be used in describing CVE instances and CWE classes is referred to as a dictionary. This dictionary is generated by the terms found in the descriptions of CVE instances and all CWE classes in the training set. Let the dictionary, denoted by $Dict$, be the set of items in the dictionary. The dictionary is empty at
the beginning, i.e., $Dict=\{\}$. Then for each text, $x_{i}$, provided either in the description of a CVE instance or in the description of a CWE class in the training set, the preprocessing step described in section (\ref{sub:Preprocessing}) is performed on it to generate a clean version of the text, $\hat{x_{i}}$. This clean version is analyzed using the three n-gram levels, i.e., 1-gram, 2-gram, and 3-gram, to generate the list of unique items $\mathbf{N}_{i}$. This
list is added to the dictionary, as described in Eq. (\ref{eq:5}).

\begin{equation}
Dict=Dict\,\cup\,\mathbf{N}_{i}\label{eq:5}
\end{equation}

Eq.  (\ref{eq:5}) computes the union between $Dict$ and $\mathbf{N}_{i}$.
After processing the text in the training set, the dictionary
$Dict$ includes all unique items that have been found in the context. Notice that any unique item appears only once in $Dict$.

The dictionary items are processed to filter out items with low frequencies (following the power law rule), i.e., a low number of occurrences in the training set. Let $f(t_{k})$ be the number of occurrences for the dictionary item $t_{k}$ in the entire training set. Then the dictionary items are filtered according to the Eq.  (\ref{eq:6}).

\begin{equation}
Dict= Dict\setminus\{t_{k}\} \& ,\, f(t_{k})<th
,\,\forall k\in{0,1,2,\text{\ldots},D-1}\label{eq:6}
\end{equation}

The threshold $th$ is the minimum number of occurrences threshold. Eq.  (\ref{eq:6}) indicates that an item $t_{k}$ is removed from the dictionary when its number of occurrences in the entire training set is less than the minimum number of occurrences threshold. This threshold is a prespecified value for the algorithm ($th=3$). The removed items are assumed to be noise and not conveying any information of interest \cite{(12)Rennie2003,(14)KHREISAT2009,(15)Egghe2000}.
% \end{itemize}

The dictionary defines the feature space, and its size determines the size of the feature vector. Let $D$ denotes the number of items in this dictionary. Each unique item in the dictionary is assigned a unique index in the feature vector. The feature vector representing CVE instance $x_{i}$ is computed as in Eq.  (\ref{eq:4}).

\begin{equation}
F_{i}[k]=\begin{cases}
1 & ,\, t_{k}\in\mathbf{N}_{i}\\
0 & ,\, otherwise
\end{cases},\,\forall k\in{0,1,2,\text{\ldots},D-1}\label{eq:4}
\end{equation}

\begin{figure}
\begin{centering}
\includegraphics[width=8cm]{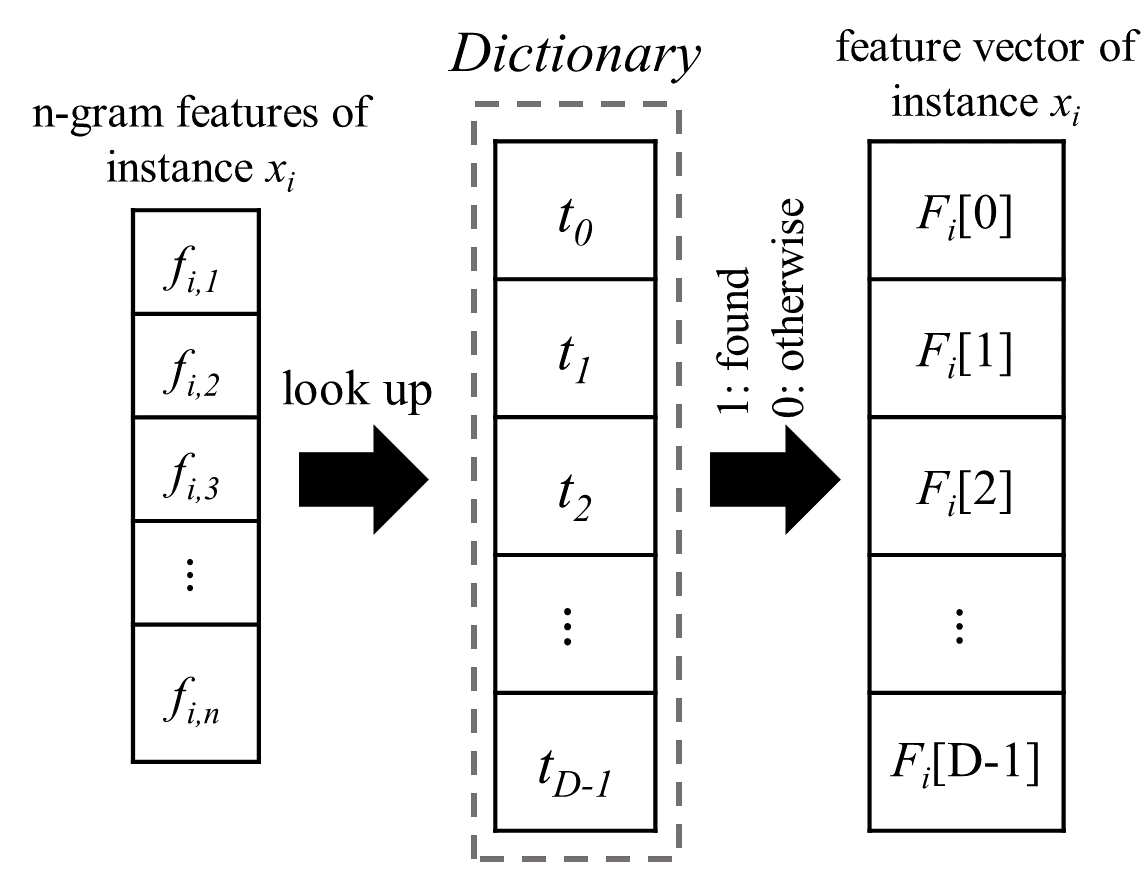}
\par\end{centering}
\caption{This shows the system design for generating feature vectors. It simply takes the n-gram feature set from each CVE description and represents (encodes) it in the dictionary space. Given the n-gram feature set for sample $x_i$, $N(x_i) = \{f_{i,1}, f_{i,2}, ..., f_{i,n}\}$ and the dictionary of features, $Dictionary = \{t_0, t_1, ...,t_{D-1}\}$, for every $t_j \in N(x_i)$, $F_{i}[j]$ is 1, otherwise 0. \label{fig:2}}

\end{figure}

The $F_{i}$ is the feature vector representing $x_{i}$. $t_{k}$ is dictionary item associated with index $k$. $\mathbf{N}_{i}$ is the list of unique items found in instance $x_{i}$.
Eq.  (\ref{eq:4}) demonstrates that the feature vector has a value of 1 at all indices corresponding to the items found in $\mathbf{N}_{i}$ and a value of 0 at the other indices. Fig. \ref{fig:2}  shows the system design for generating feature vectors.

\subsection{\label{sub:hir_dec}Hierarchical Decision-Making}
Hierarchical decision-making is the third step in the CVE classification
algorithm. This step takes the feature vectors computed in the second step (Section 3.2) and estimates the CWE classes associated with the CVE instance of interest.
% The hierarchical decision-making process uses the tree-form structure of the CWE classes provided by MITRE. In this structure, each CWE class is grouped with similar ones, creating meta-classes, which
% in turn, is grouped again until the root level is reached.
It follows a top-down approach by training a classifier per node in the CWE tree. For each node, a decision
is made using the classifier that either leads down to a different
classification node or to a leaf. This hierarchical architecture allows each classifier to focus on learning two aspects of its children: learning the details that distinguish between the classes of its children and learning the common features between the classes forming each child. This helps to simplify the design of the classifiers to reduce the computational cost, as well as improving the accuracy of the system by making each classifier to focus on its children without getting confused with other nodes.   

In the classification procedure, for each node, the neural network takes the feature vector and predicts the label of the class that is associated with a CVE instance in the next level of the tree. In this case, the neural network starts to classify CVEs to CWEs that exist at level 1. In the next level, the neural network takes the predicted CWE classes in the previous level and takes their children for the next level classification. The neural network at each node is unique and independent from the other ones at other nodes \cite{aghaei2019threatzoom}. 

According to the MITRE classification, CWE classes are not unique, which means, a CVE may belong to more than one class. Therefore, the system will miss a portion of the data if it only considers children of one selected node. In the best-case scenario, it will reach the correct class in the path, but it is unable to detect other related CWEs that may exist on the other paths since the path to the desired class is not necessarily unique. The neural network employs a multi-hot representation to label the instances to handle this multi-label problem. In this case, every instance can be represented by more than one class in training, and the neural network may output multiple predictions for each as well. 

\begin{figure}
\begin{centering}
\includegraphics[width=12cm]{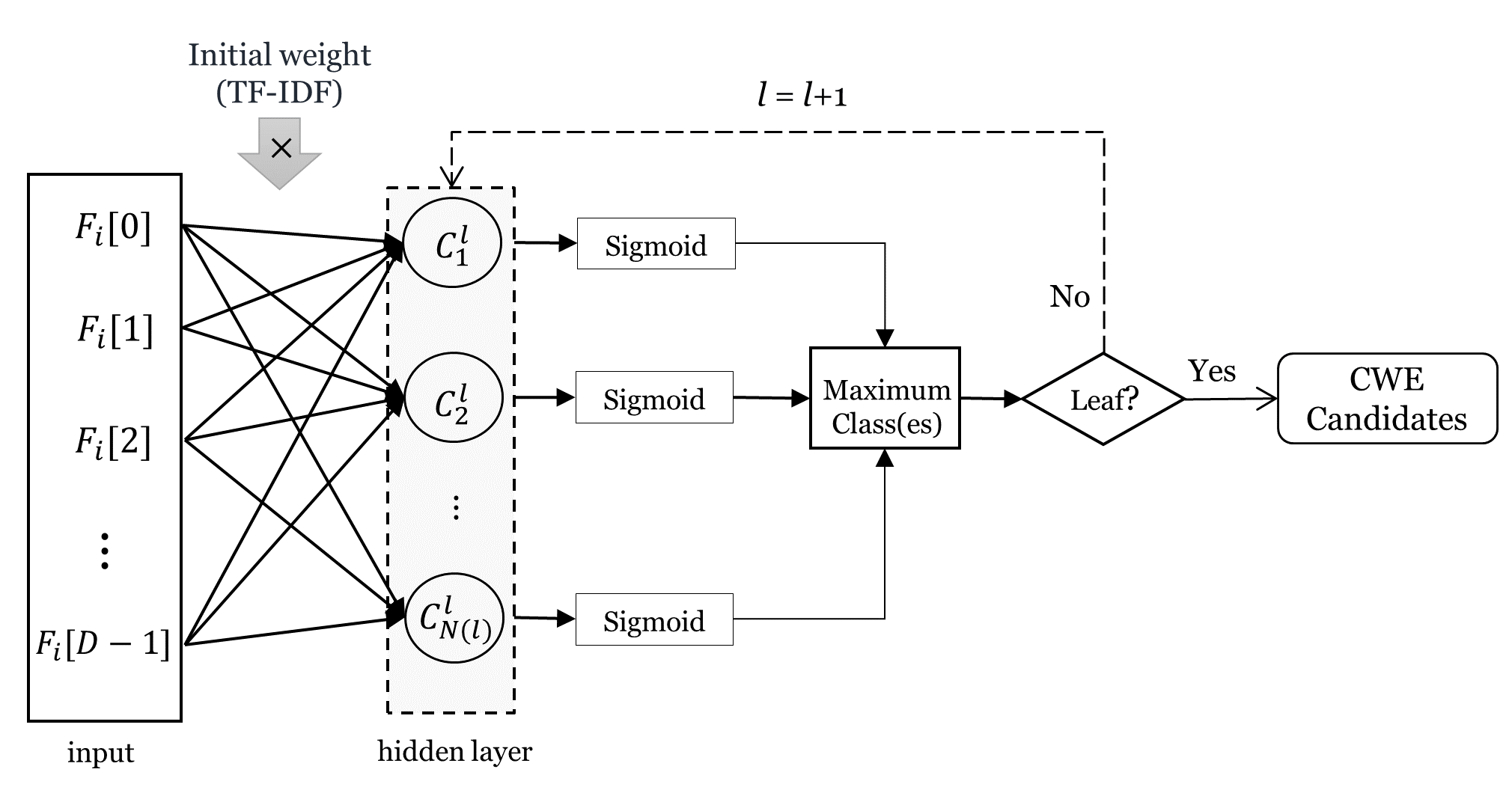}
\par\end{centering}

\caption{This shows the  hierarchical neural network design of ThreatZoom. The weight of this network is initialized by the TF-IDF score of each entry feature neuron and its corresponding class in the hidden layer. Every node in the hidden layer generates a probability score using sigmoid function and the maximum classes are assigned as the predicted classes for the input CVE. This process continues until the leaf node is reached.}\label{fig:5}
\end{figure}

Each neural network consists of one hidden layer. The number of neurons in the hidden layer equals to the number of the classes present at the level L in the CWE tree. Each neuron is fully connected to the input feature vector with no bias. The sigmoid function in every output neuron generates a probability score. If this value is higher than an imposed threshold, it means the input feature vector belongs to the class associated with that neuron. Fig. \ref{fig:5} shows the structure of the used neural network.

Let $O_{i}^{l}$
be the output of the neuron at index $i$ in hierarchy level $l$.
Then the output of this neuron before applying the Sigmoid function
is computed in eq. \ref{eq:7}.

\begin{equation}
O_{i}^{l}(x_{j})=\sum_{k=0}^{D-1}F[k].w_{ki}^{l}\label{eq:7}
\end{equation}
\\

where $x_{j}$ is the text corresponding to the input CVE instance
$j$, and $w_{ki}^{l}$ is the weight connecting the dictionary element
at index $k$ with the neuron at index $i$ at level $l$. 

The set of classes with the highest outputs at all levels is the list of candidate CWE classes output by the algorithm. 

Each neural network is trained independently from the other ones using the back-propagation procedure. The training process starts by initializing each weight with its corresponding TF-IDF value. Then the weights are iteratively adjusted to reduce the classification error. Thus, the weights of irrelevant features for the classification process keep reduced, while the weights for relevant features get increased. The obtained weights at the end of the training process are assumed to resemble the TF-IDF values that would be calculated if a large enough of training set is available. 

The classifier used in this algorithm is a single-layer neural network. This simple classifier has been employed for three main reasons: first, to reduce the computational cost by reducing the number of variables that a neural network needs to learn, second, to avoid the over-fitting problem especially when the size of the training dataset is small, and third, to allow mimicking experts thinking process when classifying CVE to CWE according to the terms mentioned in its description and their knowledge about the CWE classes. Some terms may have higher weights than others in the decision process. This can be interpreted as a score indicating the number of appearances a term occurs in the description of a CWE class, or in the descriptions of the CVE instances classified to that class, as well as, how often this term is used in other classes. This score can be estimated using the TF-IDF technique. However, only taking the TF-IDF scores into the account is error prone and is not the most effective way for two reasons: (1) computing TF-IDF values accurately requires large datasets, and (2) it is not clear what the terms that made experts do their classification are. These issues are resolved by designing a neural network that reflects TF-IDF calculations and allows it to train these values to find the ones that are important for the classification process.

\section{\label{sec:Results}Results}
The CVE-to-CWE classification algorithm is rigorously tested to quantitatively evaluate its performance and compare it with other architectures. The quantitative evaluation is performed over MITRE and NVD datasets, where both contains CVEs labeled by CWEs with a different manual approach.
The evaluation is performed by comparing the results from the
proposed algorithm with the CVE instances that have been manually associated with CWE classes by experts in MITRE or NVD, established as the ground truth.
We compute the evaluation metrics including precision, recall, and $F_{1}$ score to measure the performance of the algorithm after comparing ThreatZoom's results with the ground truth. 

\subsection{Dataset Specification}
\label{section:datset}
\begin{figure}
\centerline{\includegraphics [width=0.55\textwidth]{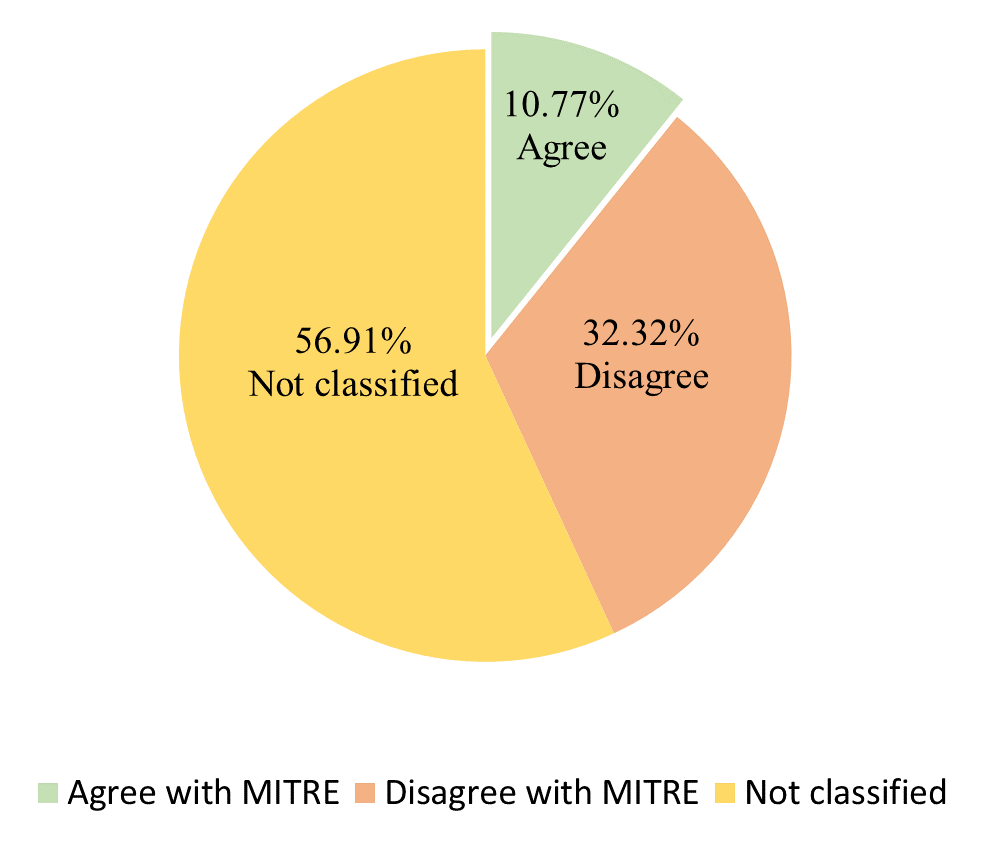}}
\caption {MITRE and NVD classification comparison}
\label{fig:MITRE-NVD_dif}
\end{figure}

There are two datasets considered in the evaluation process: MITRE dataset and NVD dataset. Each dataset consists of a group of CVE instances that have been classified to one or more CWE classes. The classification process is performed independently, so they may agree or disagree based on the different perspectives of the experts.
MITRE dataset comprises 2534 CVE instances that are classified to 364 out of 719 CWE classes.
It contains 1546 CVE instances such that, each instance is assigned to one CWE class, while each instance in the remaining CVE instances have been assigned to two or more different CWE classes.
On the other hand, NVD datasets contains more labeled CVEs compared with MITRE. However, both data repositories have some major differences. Here, w dig more deeply into the characteristic of both datasets: 
 \begin{enumerate}
 \item NVD delivers coarse-grain classification, which provides more general categories for CVEs comparing with MITRE's fine-grain classification.
 \item NVD uses the top 116 out of 719 most common CWEs in their classification, while this number for MITRE is 364.
 \item Despite MITRE, NVD does not follow the CWE hierarchy in which each CVE is classified to exactly one CWE ID.
 \item NVD does not follow MITRE's classification. Out of 2534 CVEs classified by MITRE, NVD covers only 1092 of them in their classification. Considering all the nodes in the full path MITRE classification for these 1092 CVEs, NVD's classification exists in this path only in 273 cases,  (Fig. \ref{fig:MITRE-NVD_dif}).
 \item About 40\% of the CVEs are classified to either \textit{Other} or \textit{Categorical CWE}. \textit{Other} represents an unknown CWE class, and \textit{Categorical CWE} is a superclass that covers multiple CWEs.

 \end{enumerate}
\subsection{Experiments}
A comparative evaluation experiments have been conducted to evaluate the
performance of the CVE-to-CWE classification algorithm on the labeled
CVE sets provided by MITRE and NVD. The evaluation compares the estimated candidate
list generated by the algorithm with the CWE class associated with the
CVE instance in two different levels namely coarse-grain and fine-grain which are defined as follows:
\begin{itemize}
    \item \textbf{Coarse-grain}: if ThreatZoom outputs one or a sequence of CWE classes which are not necessarily full path toward the leaf, it is considered as coarse-grain classification. In this scenario, ThreatZoom may not be able to successfully offer the entire path to the leaf but it provides few candidates for a CVE among existing CWEs. Few candidates simply mean part of the full path that is still very important to learn high-level information about the vulnerability.
    \item \textbf{Fine-grain}: if ThreatZoom successfully predicts the full path to the leaf, it is considered as fine-grain classification. Here the leaf means the original labels the datasets.
\end{itemize}

The performance of the ThreatZoom is evaluated by three standard measurements in both fine-grain and coarse-grain classification \cite{(30)serpen2018host}:
% \begin{itemize}

- \textit{Accuracy}: It computes the number of instances that are correctly classified, divided by the total number of instances. For each input CVE, the neural network offers one or more CWE candidates at each level. According to MITRE,  if the CVE label is included in the CWE candidates  it is considered as correct classification. This indicates that the algorithm provides a more detailed classification along the right path for classifying the CVE. This can be explained in part by either the CWE sub-class was created by MITRE experts after this CVE has been classified and they classified similar CVEs to this new CWE sub-class, or there is a disagreement between experts, some consider it is acceptable to stop classification, while others classified similar ones to the deeper levels. 

- \textit{Error}: It measures average number of incorrectly classified instances. Incorrect classification happens when the CVE's label is not included in CWE candidates generated by the neural network.

- \textit{Recall}:  Recall or True Positive Rate (TPR) addresses the average of correct classification per class. 

- \textit{Precision}: It represents the TPR divided by the sum of the TPR and False Rate Positive (FPR). FPR represents the rate of the instance mistakenly classified. 

- \textit{F-score}: It measures the average weighted of TPR and precision.

% \end{itemize}

Three experiments have been conducted to evaluate different aspects of the proposed design. In the first experiment, ThreatZoom has been evaluated without any modification. According to the results depicted in Table \ref{tab:tz_result}, ThreatZoom successfully results $92\%$ and $94\%$ accuracy in fine-grain and coarse-grain classification in the NVD set. In addition, it shows $75\%$ and $90\%$ classification accuracy in MITRE dataset. Although there is $17\%$ gap in accuracy of classification between NVD and MITRE, ThreatZoom shows it is able to learn short corpus and imbalance data with a promising performance, and it can perform even better if it receives more training examples from MITRE.
\begin{table}
    \centering
    \caption{ThreatZoom fine-grain and coarse-grain classification performance over MITRE and NVD datasets}
    \label{tab:tz_result}

    \fontsize{9}{9}
    \begin{tabular}{*{3}{p{.17\linewidth}|p{4.8cm}|p{4.5cm}}}% The target layout does not centre the text so we don't want \centering
      \toprule
      \centering
       & \textbf{MITRE (fine-grain, coarse grain)} &  \textbf{NVD (fine-grain, coarse grain)}\\\midrule
     
      Accuracy & 0.75, 0.90 & 0.92, 0.94 \\\midrule
      Error rate & 0.25, 0.1 & 0.08, 0.06\\\midrule
      Recall & 0.73, 0.88 & 0.90, 0.90\\\midrule
      Precision & 0.75, 0.88 & 0.91, 0.93\\\midrule
      F1-score & 0.74, 0.88 & 0.89, 0.91\\\midrule
      \bottomrule
    \end{tabular}
  \end{table}
  
\begin{figure}
\centerline{\includegraphics [width=0.6\textwidth]{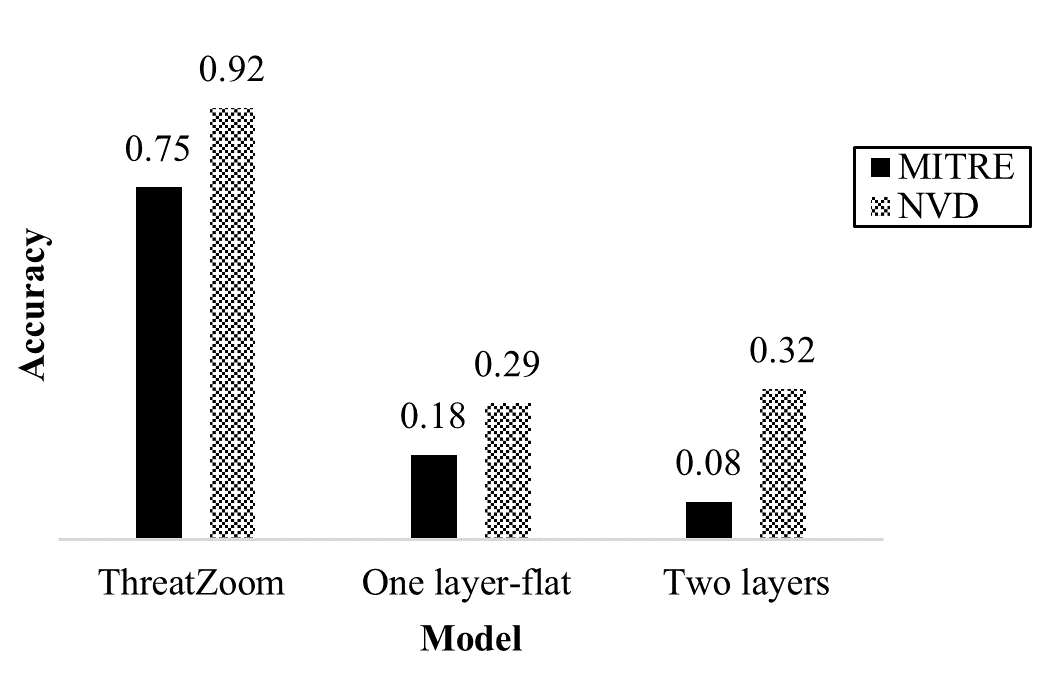}}
\caption {Comparing ThreatZoom with other models in fine-grain classification performance}
\label{fig:random_weight}
\end{figure}

 In the second experiment, the hierarchical neural network framework is replaced by a regular flat one such that the neural network considers all the classes in a one-shot classification process regardless of their hierarchy and initialized the weights randomly during the training process. 
In the third experiment, the single-layer neural network classifier is replaced by a two-layer one, and the neural network weights initialized randomly during the training process. These experiments have been conducted using MITRE and NVD datasets. For the MITRE dataset, 2131 CVEs are used for training, and 403 CVEs are used for testing. For the NVD dataset, 50000 CVEs have been used in the experiment, 40000 CVEs for training, and 10000 CVEs for testing. In each test, the classification accuracy of the testing is evaluated. In the training process, the maximum number of allowed iterations is set to 500. Fig. \ref{fig:random_weight} represents the accuracy of ThreatZoom in all three experiments. The results show that the proposed ThreatZoom approach outperforms all other approaches in classifying CVE instances to their corresponding CWE classes. The proposed ThreatZoom scores $75\%$ and $92\%$ for MITRE and NVD, respectively. The one layer-flat framework scores $18\%$ and $29\%$ for MITRE and NVD, respectively. The model with a two-layer neural network classifier scores $8\%$ and $32\%$ for MITRE and NVD, respectively.  

Similarly, ThreatZoom shows a higher performance when it has a single-layer neural network classifier, compared to the two-layer one. This can be explained by the two layers neural network has a lot more weights to learn compared to the single-layer one. This may cause an over-fitting problem when the size of the training is relatively small compared to the number of weights. In this case, the neural network learns the samples correctly in the training set while it is unable to predict correctly unseen ones. The accuracy of the two-layer neural network over the training set is $87\%$ and $92\%$ for MITRE and NVD, respectively, while it is very low on the testing set, i.e., $8\%$ and $32\%$ for MITRE and NVD, respectively. The accuracy of the two-layer neural network is high for NVD compared to MITRE that is because it has a more extensive training set, but it is still not enough to learn all the weights.   

ThreatZoom performs better when it has a hierarchical neural network architecture compared to the flat one. The flat neural network uses one classifier to learn the general and the detailed features of all the CWE classes at once. Learning general features deemphasize details to distinguish between dissimilar classes, while learning detailed features deemphasize general features to differentiate similar CWE classes, hence the confusion. On the other hand, in the hierarchical framework, each classifier learns the features that distinguish the CWE classes, which are children to its node in the hierarchical tree, only. Hence, the high performance of the proposed ThreatZoom.
% The more complicated models such as 
\section{Discussion}
In this section, we define different scenarios to point out the advantages and practicability of ThreatZoom in dealing with a variety of challenges.
\subsection{ThreatZoom and unlabeled CVEs}
ThreatZoom is tested by unlabeled CVEs, and the results are evaluated by domain experts to analyze performance consistency.
For example, CVE-2019-7632 represents "\textit{LifeSize Team, Room, Passport, and Networker 220 devices allow Authenticated Remote OS Command Injection, as demonstrated by shell meta-characters in the support/mtusize.php mtu\_size parameter. The lifesize default password for the \textit{cli} account may sometimes be used for authentication}". The proposed classification algorithm result for this CVE is the following sequence: CWE-707 $\rightarrow$ CWE-74 $\rightarrow$  CWE-77 $\rightarrow$ CWE-78. Table \ref{tab:unlabel} shows the ID and the description of the CVE and classified CWEs. The key point in this vulnerability is allowing Remote OS Command Injection. The Command Injection is a type of attack which executes arbitrary commands on the host operating system via a vulnerable application. Thus, the attacker-supplied operating system commands that are usually executed with the privileges of the vulnerable application. Regarding this definition, from the high-level perspective,  the Command Injection can be exploited under no or improper enforcement of the password, which is the \textit{lifesize default password} in this CVE. Hence, the CWE-707 clearly reflects this weakness of the system in this context. In the next level, the CWE-74 clearly addresses the \textit{injection}, which is the main action in this CVE. \textit{Command Injection} and finally the \textit{OS Command Injection} are the more specific definition of the \textit{injection} that are explained in CWE-77 and CWE-78.

\begin{table}
    \caption{Example of an unlabeled CVE (CVE-2019-7632) and neural network classification result}
    \label{tab:unlabel}
\tiny
    \fontsize{6}{6}
    \begin{tabular}{*{2}{p{.3\linewidth}|p{8.5cm}}}% The target layout does not centre the text so we don't want \centering
    
      \toprule% nicer rules courtesy of booktabs - but then we need to drop the verticals 
      \textbf{ID} &  \textbf{Description} \\\midrule% Note that there is no & before the first column - & only comes between columns so if you define n columns, you can have at most n-1 & symbols in any row
     \fontsize{5}{5}
      CVE-2019-7632 & LifeSize Team, Room, Passport, and Networker 220 devices allow Authenticated Remote OS Command Injection, as demonstrated by shell meta-characters in the support/mtusize.php mtu\_size parameter. The lifesize default password for the cli account may sometimes be used for authentication.\\\midrule% the p{} setting automatically lets these be multi-line - we don't want multiple rows on top of that and this is simpler as TeX does the hard work for us

      CWE-707: Improper Enforcement of Message or Data Structure & The software does not enforce or incorrectly enforces that structured messages or data are well-formed before being read from an upstream component or sent to a downstream component.\\\midrule
          		
    CWE-74: Injection & The software constructs all or part of a command, data structure, or record using externally-influenced input from an upstream component, but it does not neutralize or incorrectly neutralizes special elements that could modify how it is parsed or interpreted when it is sent to a downstream component.\\\midrule
    
    CWE-77: Command Injection & The software constructs all or part of a command using externally-influenced input from an upstream component, but it does not neutralize or incorrectly neutralizes special elements that could modify the intended command when it is sent to a downstream component. \\\midrule
    		
    CWE-78: OS Command Injection & The software constructs all or part of an OS command using externally-influenced input from an upstream component, but it does not neutralize or incorrectly neutralizes special elements that could modify the intended OS command when it is sent to a downstream component.\\\midrule
    %   \bottomrule
    \end{tabular}
  \end{table}

\subsection{More fine-grain classification by ThreatZoom}
\begin{figure}
\centerline{\includegraphics [width=0.7\textwidth]{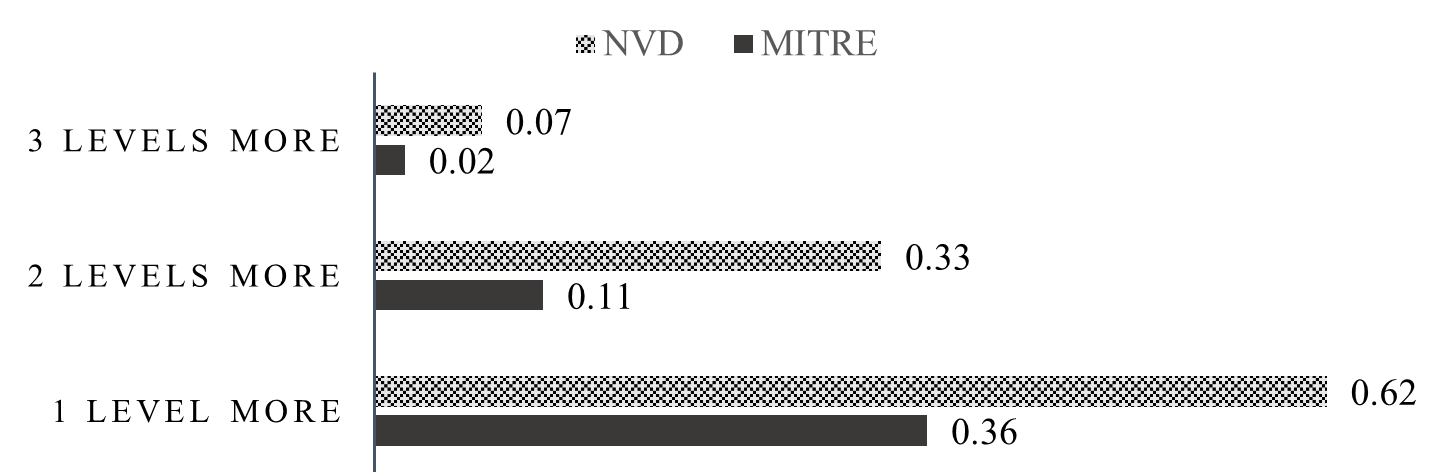}}
\caption {more fine-grain classification of ThreatZoom over MITRE and NVD classification}
\label{fig:betterthan}
\end{figure}

Reported results show not only ThreatZoom performs as good as the MITRE and NVD engineers, who are well-trained for this task, but also it accumulates their experience to offer even more fine-grain CWEs. Fig. \ref{fig:betterthan} shows our approach obtains more fine-grain classification for about $47\%$ and $95\%$ of those CVEs that were correctly matched by MITRE and NVD, respectively. To validate these results, we used our domain expertise to manually inspect 100 randomly selected CVEs that received fine-grain classification by our tool. For example, consider "\textbf{CVE-2001-1386}: WFTPD 3.00 allows remote attackers to read arbitrary files by uploading a (link) file that ends in a '.lnk.' extension, which bypasses WFTPD's check for a ".lnk" extension", where WFTPD is an FTP server for Windows systems. The CVE implies that this server contains a directory traversal vulnerability, which may allow users to upload files with a name extension '.lnk'. ThreatZoom classifies this vulnerability to fine-grain class, which is "\textbf{CWE-65 (Windows Hard Link)}: The software, when opening a file or directory, does not sufficiently handle when the name is associated with a hard link to a target that is outside of the intended control sphere. This could allow an attacker to cause the software to operate on unauthorized files". It clearly addresses the \textit{windows-based software} flaw that\textit{ does not properly validate the name} of the uploaded file that causes unauthorized operation over the data. This CWE is accurate and more specific than the MITRE classification which is "\textbf{CWE-59 (Improper Link Resolution Before File Access ('Link Following'))}: The software attempts to access a file based on the filename, but it does not properly prevent that filename from identifying a link or shortcut that resolves to an unintended resource".

\section{Conclusion and Future Work}
\label{sec:Conclusion}

There are continuous efforts by various organizations to provide a quality CVE-to-CWE classification to enable a deep understanding of vulnerabilities and facilitate proactive threat defense. However, all existing efforts are mainly manual and thereby do not cope with the rate of CVE generation. This paper presents an automated technique for fine-grain classification of CVE instances to CWE classes. To this end, we discuss the existing challenges and shortcomings and introduce ThreatZoom as a novel blend of text mining techniques and neural networks in order to overcome problem. To the best of our knowledge, ThreatZoom is the first automated system that addresses existing challenges and limitations of classifying CVEs to CWEs.
Our approach is rigorously evaluated using datasets provided by MITRE and NVD. The accuracy of classifying CVE instances to their correct CWE classes using the MITRE dataset is $75\%$ and $90\%$ for fine-grain and coarse-grain classification, respectively, and $92\%$ for NVD dataset.
In many cases, our approach obtains a more in-depth fine-grain classification for $36\%$ for MITRE, and $62\%$ for NVD, as verified by domain experts. 
This tool is already being developed to leverage word embedding, and deep neural networks to (1) better classify CVEs to CWEs following MITRE standards, (2) extract threat actions out of CTI reports, (3) automatically calculate threat severity scores, and (4) find the potential course of action mitigations and detection, accordingly.
\section*{Acknowledgement}
This research was supported in part by the Office of Naval Research (ONR) and National Science Foundation (NSF). Any opinions, findings, conclusions or recommendations stated in this material
are those of the authors and do not necessarily reflect the
views of the funding sources.
\bibliographystyle{splncs04}
\bibliography{SECURECOMM20}

\end{document}